# Anatomy of a superorganism – structure and growth dynamics of army ant bivouacs

Thomas Bochynek[1,2,†], Florian Schiffers[2], André Aichert[4], Oliver Cossairt[2], Simon Garnier[5], and Michael Rubenstein[1,2,6,*]

Author affiliations:
1 -- McCormick School of Engineering, Center for Robotics and Biosystems, Northwestern University, Evanston, IL 60201, USA
2 -- Computer Science, Northwestern University, Evanston, IL 60201, USA
† -- current affiliation: Department of Physics, Emory University, Atlanta, GA 30322, USA
4 -- Siemens Healthineers, Digital Technology & Innovation, 91052 Erlangen, Germany
5 -- Department of Biological Sciences, New Jersey Institute of Technology, Newark, NJ 07102, USA
6 -- Mechanical Engineering, Northwestern University, Evanston, IL 60201, USA

*Corresponding author: Michael Rubenstein, 2145 Sheridan Road, B224, Evanston, IL 60208
Phone: 847-467-4685
Email: rubenstein@northwestern.edu

**This PDF file includes:**
  Main Text
  Figures 1 to 3

## Abstract

Beyond unicellular and multicellular organisms, there is a third type of structural complexity in living animals: that of the mechanical self-assembly of groups of distinct multicellular organisms into dynamical, functional structures. One of the most striking examples of such structures is the army ant bivouac, a nest which self-assembles solely from the interconnected bodies of hundreds of thousands of individuals. These bivouacs are difficult to study because they rapidly disassemble when disturbed, and hence little is known about the structure and rules that individuals follow during their formation. Here we use a custom-built Computed Tomography scanner to investigate the details of the internal structure and growth process of army ant bivouacs. We show that bivouacs are heterogeneous structures, which throughout their growth maintain a thick shell surrounding a less dense interior that contains empty spaces akin to nest chambers. We find that ants within the bivouac do not carry more than approximately eight times their weight regardless of the size of the structure or their position within it. This observation suggests that bivouac size is not limited by physical constraints of the ants' morphology. This study brings us closer to understanding the rules used by individuals to govern the formation of these exceptional superorganismal structures, and provides insight into how to create engineered self-assembling systems with, for instance, swarms of robots or active matter.

**Significance Statement**

Army ant bivouacs are one of the most complex self-assemblies in the animal kingdom, and notoriously difficult to study. We observe their assembly by using a custom field-worthy CT scanner, and measure growth dynamics and physical properties. We find that the bivouac interior contains empty spaces akin to nest chambers and is surrounded by a dense outer layer of ants. Physical stresses experienced by individuals are distributed equally amongst the ants, independently of their location within the bivouac, suggesting that army ants create a structure that has no theoretical upper size limit.

## Introduction

Self-assembly is a major driving force in the evolution of complexity and adaptability in biological systems (1, 2). The formation of complex proteins from amino acids and of cell membranes from lipid molecules marked the transition from chemical to biological activity (3). Subsequently, the assembly of individual cells into multicellular organisms permitted the emergence of an incredible diversity of life forms from, essentially, shared building principles and common cellular bases.

A third level of biological self-assembly is less often considered: the formation of organized patterns from the collective activity of multicellular organisms. These patterns include, for instance, flocks of birds, schools of fish, and herds of mammals which, unlike the cells and molecules in our bodies, are not bound by chemico-physical interactions but by less tangible social ones (4). A special case of self-assembly is the formation of three-dimensional structures with defined function from physically interconnected bodies, found in certain social insect species. These mechanical assemblies represent a culminating point in the evolution of the extended morphology of these colonial superorganisms (5). Some of the best-known examples include the pulling chains which arboreal Australian weaver ants form and use to bend leaves into living nests (6), the rafts on which fire ants escape rising floodwaters (7), the bridges which army ants use to cross gaps in their foraging trails (8), and the large, tightly packed clusters that honeybee workers form to protect their queen during migrations (9, 10).

All these examples of mechanical assemblies pale in comparison to the complex structure of the bivouacs of *Eciton* army ants. These bivouacs are the temporary structures that these nomadic social insects build out of their own bodies between episodes of migration, and which they use as a nest and a staging ground for their extensive foraging raids (11). Often conical in shape and usually suspended from a supporting structure like a fallen tree log (Fig. 1) or within a natural cavity (11), the bivouac of a mature colony can be formed of hundreds of thousands of interlocked workers (12, 13). It has internal empty spaces that likely function as chambers to hold and protect the queen and brood (14), and its structure adjusts itself to temperature variations (15). During their nomadic phase, these near-blind ants disassemble the entire structure every night, and within the span of a few hours reassemble it hundreds of meters away, without any apparent central coordination (11). The specifics of the self-assembly process, as well as the internal organization of army ant bivouacs, have been largely unknown: army ants react ferociously to physical disturbance (see Movie S4), and attempts at directly observing the inside of a bivouac inevitably lead to quick and drastic changes of its structure. For this reason, previous work on the topic has largely been limited to qualitative field observations.

Using a custom-built X-ray Computed Tomography (CT) scanner, we present a quantitative study of the bivouac's external and internal structural properties, with the aim of better understanding the behaviors underlying its self-assembly process. We reveal that bivouacs are complex and heterogeneous structures with a partially empty interior surrounded by a well-defined outer shell of approximately 1.4 cm thickness. This outer shell is maintained as additional ants join the structure and the bivouac grows. The load carried by individual ants appears to be identical throughout the bivouac independent of an ant's location, implying that the size of the bivouac is not limited by the load-supporting capacity of individuals. The insights into the structure and the functioning of this remarkable biological system can also provide valuable inspiration to the fields of swarm robotics and active matter, both of which are interested in the self-assembly of artificial agents into complex, functional structures resembling those produced by nature (16–18).

## Results and Discussion

In this study, we focused on investigating the structural properties and the internal structure of the bivouacs as they grow over several orders of magnitude during the self-assembly process. We created 3D reconstructions of the self-assembly process of bivouacs from collected colonies, First, we investigated how the density of the bivouacs changed as they grew in volume and number of ants. To estimate the number of ants in a bivouac, we manually counted ants in sparsely populated bivouac areas, integrated the X-ray absorption values from the corresponding CT scan for the same areas, and derived an absorption-to-ant-count scaling factor. To quantify bivouac volume, we fitted a bounding hull around the bivouac (Fig. S2) and computed its volume. We found that the bivouac volume grew faster than the number of ants, resulting in a superlinear relationship between the two measurements (Fig. 3A). Two hypotheses could explain this reduction of the bivouac density (i.e., the number of ants per unit of volume) with increasing mass (i.e., the number of ants): 1) if bivouacs are homogeneous structures, the density reduction could be the result of ants increasing the distance between individuals as the bivouac grows, or 2) if bivouacs are heterogeneous, the density reduction might result from the emergence of regions of differing morphology which scale differently as the bivouac grows (e.g., bivouacs might be composed of a dense central core to which less-dense outer layers are added).

To investigate the source of this relationship, we inspected bivouacs for structural heterogeneities. In particular, we measured how the bivouac density changed as a function of the distance to the bivouac's outer boundary. We found that, on average, a bivouac could be separated into two regions of density: an outer shell approximately 1.4 cm thick, and an interior volume approximately half as dense (see Fig. 3B). Shell thickness increases slightly but significantly with the size of the bivouac, and shell density decreases with the distance to the outer boundary in smaller bivouacs while it appears constant in large, presumably stable bivouacs. However, the density of the interior volume is constant on average.

Together, these results suggest that the observed positive allometric scaling between bivouac volume and number of ants in the bivouac results from an uneven scaling between a dense outer layer of fixed thickness (the volume of which scales approximately with the second power of the number of ants) and a less-dense core volume (which scales with the third power).

We hypothesize that the bivouac grows by the addition of ants to the outside of this shell, while former surface regions that now form part of the interior become less dense, presumably due to ants inside the bivouac detaching once they are no longer needed to maintain the integrity of the structure. Indeed, we find that only 23.3% of all bivouac voxels that were occupied at any point during the growth process remained occupied until the end of the experiment (Table S1). This observation is consistent with results from the bridge-building behavior of army ants, where ants that are no longer necessary for the functioning of the structure detach themselves and re-join the rest of the worker pool (8).

Finally, we investigated how the observed bivouacs distribute and support the weight throughout their structure to see what relative stress individual ants must support as a function of their position in the bivouac. We were interested in determining whether ants near the top of the bivouac were subjected to greater stress than ants attached below them. If that were the case, it would suggest that the mechanical stress on the ants located in the upper region of the bivouac should increase as the size of the bivouac grows, therefore imposing a physical limit to the maximum size of the bivouac. On the other hand, if the stress at the top was not significantly greater than that in the layers below, this would instead suggest that the load-supporting capacity of individual ants does not limit the maximum size of these bivouacs. To answer this question, we considered horizontal slices of thickness 0.76 cm (the approximate average length of a worker (19)) and compared the slice's total mass (estimated via proxy of the summed X-ray absorption of the slice) with the mass of the rest of the bivouac below it. Dividing the latter by the former gives an estimate for the load supported by each ant within the slice. We found that the load supported by an individual ant is close to constant at approximately eight times the mass of an ant in the upper two thirds of the bivouac, and much lower in the bottom third (Fig. 3C, Fig. S7). This result is consistent with the hypothesis that the self-assembly process inherently enforces an upper limit to the load supported by the ants, and implies that the maximum size of a bivouac is not limited by the load-supporting capacity of individual ants but only by the number of available ants. Thus the construction rules of the army ant bivouac are not a limiting factor to the growth of the colony.

Using our custom-made CT scanner, we were able to obtain a detailed, quantifiable description of the growth and internal structure of army ant bivouacs. Our approach allowed us to determine that bivouacs grow most likely by the addition of ants to the periphery while some workers in the interior relocate, presumably to make space for brood and ants that are not a structural part of the bivouac assembly. Ultimately, the process results in a conic outer shell and a heterogeneous interior with pockets of high and low density without an apparent regular distribution.

Despite being in a state of constant reconfiguration, these bivouacs maintain distinguishable morphological regions, achieve an equal distribution of carried loads, and appear to circumvent upper size limits. This is particularly remarkable considering that individual workers attach and detach from existing structures purely based on local information (1) and without apparent knowledge of the total bivouac size or their absolute location within the bivouac. Previous studies on less complex army ant self-assemblages such as pothole "plugs" (20), bridges (8, 20, 21), and "safety barriers" (22) suggest that a small set of individual behavioral rules may explain the collective construction behavior of these social insects (23). Here we provide the initial research for identifying the individual rules underlying the growth and reorganization of army ant bivouacs, a much larger and much more complex structure. In the future, this knowledge may allow researchers to use army ant self-assembly as a reference model for designing large-scale artificial self-assembling systems in the fields of swarm robotics, self-repairing materials, and active matter.

## Materials and Methods

### Experimental setup

The Computed tomography (CT) scanner consisted of a Varex Imaging M-113T X-ray tube (Fig. 2B) and a Varex Imaging XRD 4343 Digital X-ray detector (Fig. 2D). Power to the X-ray tube was provided by a Spellman PMX 5kW X-ray generator. The imaging container (Fig. 2C) was mounted on a 3D-printed rotation stage driven via a 1000:1 Micro Metal Gearmotor HP 6V located off-center and connected to the stage rim via gears. The rotation stage, X-ray generator, X-ray tube, and X-ray detector were controlled and synchronized via an Arduino Uno microprocessor. Ants could access the imaging container from the collection container (Fig. 2A) through clear silicone tubing of 2.5 cm inner diameter. All electronic parts were mounted inside a rectangular cabinet of dimensions 183 * 61 * 70 cm (length * width * height), built from 2.5 cm wide 80/20® T-Slotted Aluminum extrusion rails. The cabinet outside was clad in overlapping radiation-shielding panels (composed of a lead sheet of 1.6 mm thickness laminated onto a 1.6 mm steel sheet, purchased from Radiation Protection Products Inc). Experimenter access to the rotation stage was given via a hatch on the cabinet side. Openings for cables and silicone tubing were covered in additional U-shaped shielding to prevent radiation leakage. See Movie S4 for a depiction of the scanner and its components.

### Experiment details

We installed the custom CT scanner in the laboratories of the Smithsonian Tropical Research Institute field station on Barro Colorado Island in Panama, where we collected five near-complete *Eciton burchellii* army ant colonies with a vacuum-powered aspirator (see Movie S4). Colonies were released after 24-36 hours, during which the laboratory doors were left open to provide colonies the same temperature and humidity conditions as in their surrounding natural habitat.

From the collection container colonies had access via silicone tubing to a cylindrical imaging container of 30 cm height and diameter inside the CT scanner (see Fig. 2 and Movie S4 for setup details). A sealed passive rotation joint connected the rotating imaging container to the stationary silicon tube. The imaging container lid was lined with a cork disk of 25 cm diameter, from which the ants readily formed hanging bivouacs.

We began imaging once ants had assembled a thick chain from their bodies, and recorded 3D reconstructions in approximately 20-minute intervals afterwards. Once the approximate size of the bivouac stopped changing in subsequent scans (as determined by visual inspection of the recorded X-ray projections), we used the aspirator to relocate the ants from the imaging container to the temporarily disconnected collection container. Re-establishing the connection between the two containers restarted the self-assembly process.

A single scan took less than two minutes to complete. From the five collected colonies, we observed a total of 25 bivouacs (mean = 5 bivouacs per collected colony, SD = 1.22) throughout their construction, gathering 3 to 15 (mean = 8.24, SD = 3.14) scans for each bivouac, obtaining a total of 196 scans. The resulting reconstructions clearly show stationary ants within the bivouac (Fig. S1), while ants in motion moved quickly relative to the scanning speed and were not resolved.

Bivouacs formed inside the imaging container had a cone-like shape, consistent with the most common shape of the free-hanging bivouacs found in the field (11) (Fig. 2E,H, Movies S1 and S3). Our 3D reconstructions showed the distribution of ants and internal empty spaces within the bivouacs and even allowed for identification of the workers (Fig. S1C) and, when present, the queen ($N = 2$) (Fig. S1B). The latter was located near the top of the bivouac, consistent with previous observations (11). Our time series scans (see Movie S3) showed bivouacs growing from chains early in the process into a single, progressively expanding 3D structure (Fig. 2F-H) with a internal structure containing empty spaces (Fig. 2I-K).

The colonies we used in the analysis did not include significant amounts of brood in the bivouac. The bivouac shown in Movie S1, which contains a column of brood, was hence not used in the analysis.

**X-ray projection acquisition and 3D image reconstruction**

X-ray projections were taken while the imaging container rotated in a continuous motion. The high voltage generator required power cycling every 30 seconds which led to a down-time of 4 seconds during which the rotation was interrupted. A full imaging cycle was completed in approximately 120 seconds, which required three such power cycles. An imaging cycle consisted of approximately 450 2d projections recorded at a rate of 4.5 projections per second. Each projection was generated by an X-ray burst of 10.6 ms length at 40 keV energy and 125 mA current. The active area of the detector measured 432 * 432 mm, detected images had dimensions of 2880 * 2880 pixels with a 150 μm pixel pitch and 16-bit image information. The volume of a voxel in the 3D reconstructions was 0.058 mm$^3$. The details of the reconstruction process are given in the Supplementary information.

**Data preparation**

Raw 3D volumes contained reconstruction artifacts, e.g. low-absorption voxel values from moving ants. We applied an identical threshold to all reconstructions that set voxels with intensities $I \leq 50$ to $I = 0$. Based on visual inspections, this threshold value retained identifiable ant reconstructions but removed values of indeterminable origin.

**Calculation of bivouac bounding hull and volume**

We calculated bivouac volumes via a bounding hull placed around the bivouac (see Fig. S2). The hull was created from the ant-locations identified with the thresholding operation described above in a multi-step process: potential noise in the form of small, disconnected locations were removed via a one-time erosion and subsequent dilation with a 1-pixel radius circular kernel by using the functions *erode* and *dilate* from *OpenCV* 3.4.1 (24), applied to all slices within a bivouac reconstruction. To connect the curated ant locations into a hull, we applied a seven-fold dilation with a circular kernel of radius $r = 8$ pixels, and a subsequent seven-fold erosion with the same kernel. To smooth steps between hull locations in subsequent horizontal bivouac slices, we performed an additional one-time dilation and erosion of the hull using a 3D spherical kernel of radius $r = 5$ pixels. We filled empty spaces in the resulting hulls using the function *ndimage.morphology.binary_fill_holes* in *scipy* 1.2.1 (25). In order to remove small ant-chains that were not part of the main body of the bivouac, we identified all connected hull components in each slice, and only retained the largest one. The volume of the thus created bivouac bounding hull was then calculated by summation of the hull areas across the stack of all horizontal slices. An example of the resulting hull is shown in Fig. S2.

**Approximation of the number of ants in a bivouac**

To calculate the approximate number of ants contained in each bivouac, we derived a constant that converted the X-ray absorption values of the 3D reconstructions to ant counts. For this, we manually inspected sparsely populated bivouac subvolumes and counted the ants present in them. We then summed the X-ray absorption values of these regions, and divided it by the number of counted ants. We derived the conversion constant of $A_{single\_ant}$ = *19860*. In this way, we derived the approximate ant counts in the portion of 3D reconstructions contained within the bounding hull. We note that this proxy may not fully reflect the exact polymorphic worker size distribution within the bivouac.

**Scaling relationship between bivouac volume and number of ants inside the bivouac**

To obtain a power law function fit for the scaling relationship between the number of ants in a bivouac $N_A$ and the bivouac volume $V_B$ shown in Fig. 3A, we used the function *optimize.curve_fit* from *scipy* 1.2.1 implemented in *Python* 3.7.3 (26). The obtained power law fit was $V_B$ = *-9.33 * $N_A^{1.035}$* (dashed line in Fig. 3A), suggesting that bivouacs become less dense as they grow larger. To calculate the 95% confidence intervals for the fitted parameters by bootstrapping, we fitted the function to a random subsample of collected colonies (taken with replacement) of the dataset with the same size in $N = 10^6$ iterations. The grey area shown in Fig. 3A gives the 95% confidence region for the power law fit parameters: [1.011, 1.131] for the power law exponent and [-10.401, -8.892] for the constant).

**Horizontal structural heterogeneity: bivouac shell and interior**

In order to determine whether or not bivouacs have a heterogeneous internal structure, we calculated the local bivouac occupancy ratio (i.e. the ratio of occupied to empty voxels) as a function of distance to the outer bivouac border. We determined the distance of each voxel to the bivouac outside via an iterative voxel-wise erosion of the bivouac volume implemented using the function *ndimage.binary_erosion* from *scipy* 1.2.1 applied to the binarized 3D reconstruction (binarization thresholding is outlined above in the section Data preparation). We excluded from the analysis all bivouacs that were composed only of narrow chains with little internal volume (usually the earliest scans in the time series) and those that did not have a single, conically-shaped bivouac structure (i.e. those that consisted of several conjoined conical structures). This selection process resulted in the exclusion of 49 out of the 196 scans available, the following analysis was performed on the remaining 147 scans. Additionally, we excluded the first 10 data points representing the outmost bivouac layer from the analysis: because the hull-making algorithm attaches the hull edge to the outermost ant locations, ant density is overrepresented on the immediate hull periphery and results in the spike-artefact seen in Fig. 3B between $x = 0$ and $x = 0.2$.

The resulting data (shown in Fig. 3B) suggests two regions of different occupancy ratios: a denser, peripheral region of approximately 1.4 cm thickness on average (hereafter referred to as the shell), and a less dense interior (hereafter referred to as the core).

In the following, we used a segmentation analysis to determine the breakpoint between these two regions. We fitted a weighted segmented regression model with the occupancy ratio as the dependent variable and the distance to the bivouac's exterior as the independent variable, using the segmented package (version 1.2-0) (27) in *R* version 4.0.2 (28). Because the number of voxels used in the calculation of the occupancy ratio was higher towards the bivouac exterior, each occupancy ratio data point was weighted by the number of voxels that were considered during its calculation. The location of the breakpoint between the shell and the core segments corresponds to the thickness of the shell. The intercept of each segment corresponds to the maximum occupancy ratio of each section of the bivouac while the slope is an indication of how constant the occupancy ratio is within each section.

In order to investigate the effect of the bivouac size on the thickness of the shell, we fitted a linear mixed-effects model with the log-transformed breakpoint location as the dependent variable, bivouac size as the independent variable (rescaled using the root mean square of the data) and replicate number and colony identity as nested random effects, using the lmerTest package (version 3.1-2) (29) in *R*. The model fit was assessed by visual inspection of the diagnostics plots as produced by the *plot_model* function in the *sjPlot* package (version 2.8.4) (30) in *R*. The result of the model fit is presented in Table S2 and Fig. S3. It shows the shell thickness to be between 1.07 and 1.72 cm on average (see confidence interval of model slope in Table S2). It also shows a significant positive effect of the bivouac size on the shell thickness ($p = 0.015$).

In order to determine the effect of the bivouac size on the maximum density of the bivouac shell and core, we fitted a linear mixed-effects model with the intercept of the segmented regressions as the dependent variable, bivouac size (rescaled using the root mean square of the data) and bivouac section as the independent variables and replicate number and colony identity as nested random effects. A comparison of this model with a model including an interaction term between bivouac size and bivouac section showed that the interaction term did not significantly improve the fit and it was therefore not considered in this analysis. The model fit was assessed by visual inspection of the diagnostics plots as before. The result of the model fit is presented in Table S3 and Fig. S4. It shows that the maximum density of the shell is significantly greater than that of the core ($p < 0.001$) and that the bivouac size has a significant negative effect on the maximum density of both the core and the shell ($p = 0.019$).

Finally, we determined the effect of the bivouac size on the uniformity of the density of the bivouac shell and core by fitting a linear mixed-effects model with the slope of the segmented regressions as the dependent variable, bivouac size (rescaled using the root mean square of the data) and bivouac section as the independent variables and replicate number and colony identity as nested random effects. A comparison of this model with a model including an interaction term between bivouac size and bivouac section showed that the interaction term did significantly improve the fit and it was therefore also included in this analysis. The model fit was assessed by visual inspection of the diagnostics plots as before. The result of the model fit is presented in Table S4 and Fig. S5. It shows that, while the density of the core of the bivouac is constant on average, that of the shell decreases significantly with the distance to the bivouac edge ($p < 0.001$). That effect is affected significantly by the bivouac size in the shell section only, but not in the core (interaction term, $p < 0.001$): in small bivouacs, the density of the shell decreases quickly with the distance to the envelope, while it remains constant in large bivouacs.

**Load carriage calculation and the constraining influence of the experimental setup**

The inset in Fig. 3C visualizes the calculation of the load carriage via a sliding window of thickness $t = 0.76$ cm (indicated by the red rectangle), by dividing the summed mass of the bivouac below this window ($W_b$) by the integrated mass within the window ($W_d$). The window width is not shown to scale. Subscript *d* refers to the distance between window center and bivouac top, and subscript *b* refers to the distance from the window bottom (i.e., $d + 1/2 * t$) to the bivouac bottom.

The dimensions of the imaging container inherently impose constraints on the maximum dimensions that bivouacs can obtain. The major constraining factor is the diameter of the cork disk mounted under the lid of the imaging container, which limits the potential diameter of the bivouacs suspended from it. Fig. S6 shows that approximately 60% of all bivouacs grow to the size in which they are constrained by the cork disk diameter in the upper parts of the bivouac. To investigate whether or not this constraint causes the increase in load carriage shown in the upper bivouac ranges in Fig. 3C, we produce the same plot for a subset of $N$ = 25 bivouacs that do not encounter the edge of the cork disc. Fig. S7 shows that in these bivouacs, the increase of carried load observed in Fig. 3C is absent. Likewise, the fraction of all bivouacs that are constrained by the diameter of the cork disk decreases sharply at a distance to the bivouac top $d$ < 5 cm, which coincides with the region of increased carried load shown in Fig. 3C. Both figures suggest the increase in carried load per individual shown in Fig. 3C is imposed by spatial constraints of the experimental setup, and that in spatially unconstrained bivouacs load carriage remains fixed throughout most of the bivouac at approximately 8 times the weight of an individual.

**Internal reconfiguration analysis**

To estimate what fraction of the bivouac remains occupied throughout the experiment duration, we inspected every voxel throughout the reconstructions of a given bivouac time series. If a voxel at any given time contained an ant, we tested if it remained occupied throughout all subsequent time steps. If it did, we counted it towards the number of continuously occupied voxels. The resulting data is shown in Table S1. Of all voxels inspected in this way, only 23.3% remained occupied.

## Acknowledgments

We thank the Northwestern Memorial Hospital for their help in the preparation of our CT scanner equipment, the administrative and field staff at the Smithsonian Tropical Research Institute for their support with our field work, and Siemens Healthineers AG (Klaus Engel and Frank Dennerlein for the cinematic rendering of our 3D reconstructions, and Tobias Wuerfel for his feedback on CT hardware and software design).

**Figures and Tables**

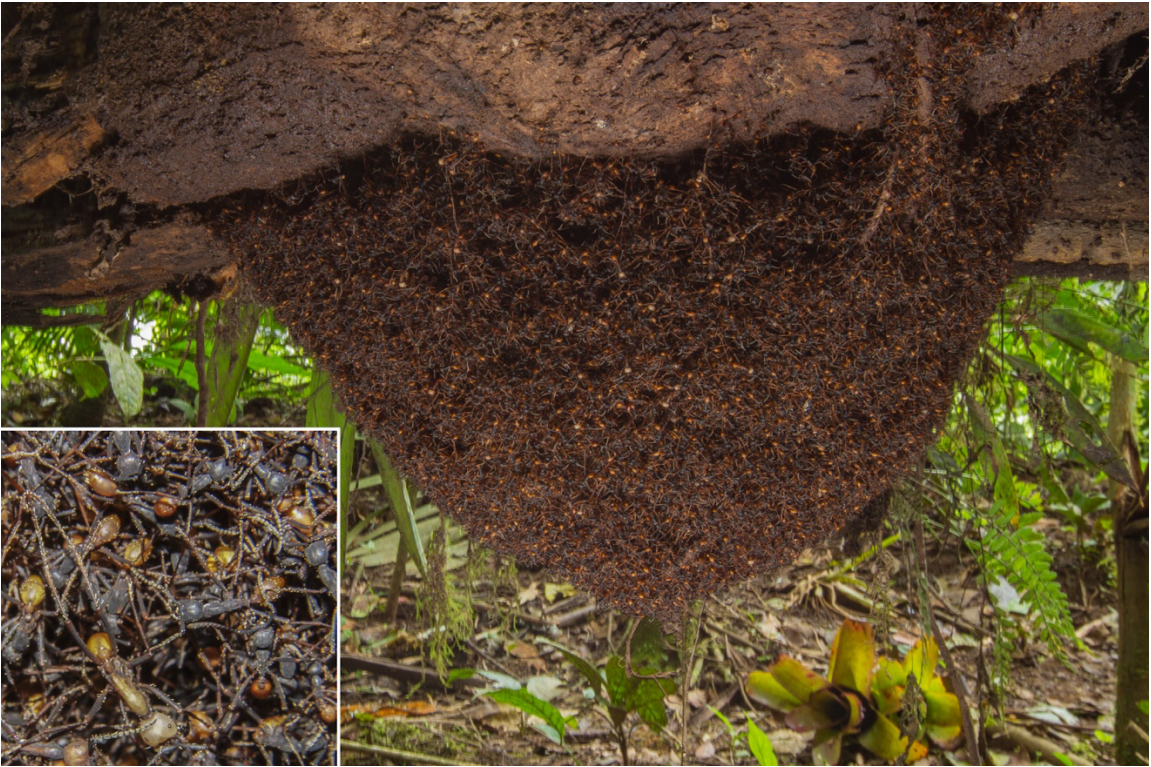

**Figure 1.** Army ant bivouac in nature. An example of an army ant bivouac in a natural environment. Inset: interconnected individuals forming a bivouac. Images courtesy of Daniel Kronauer (background reproduced with permission from (22)).

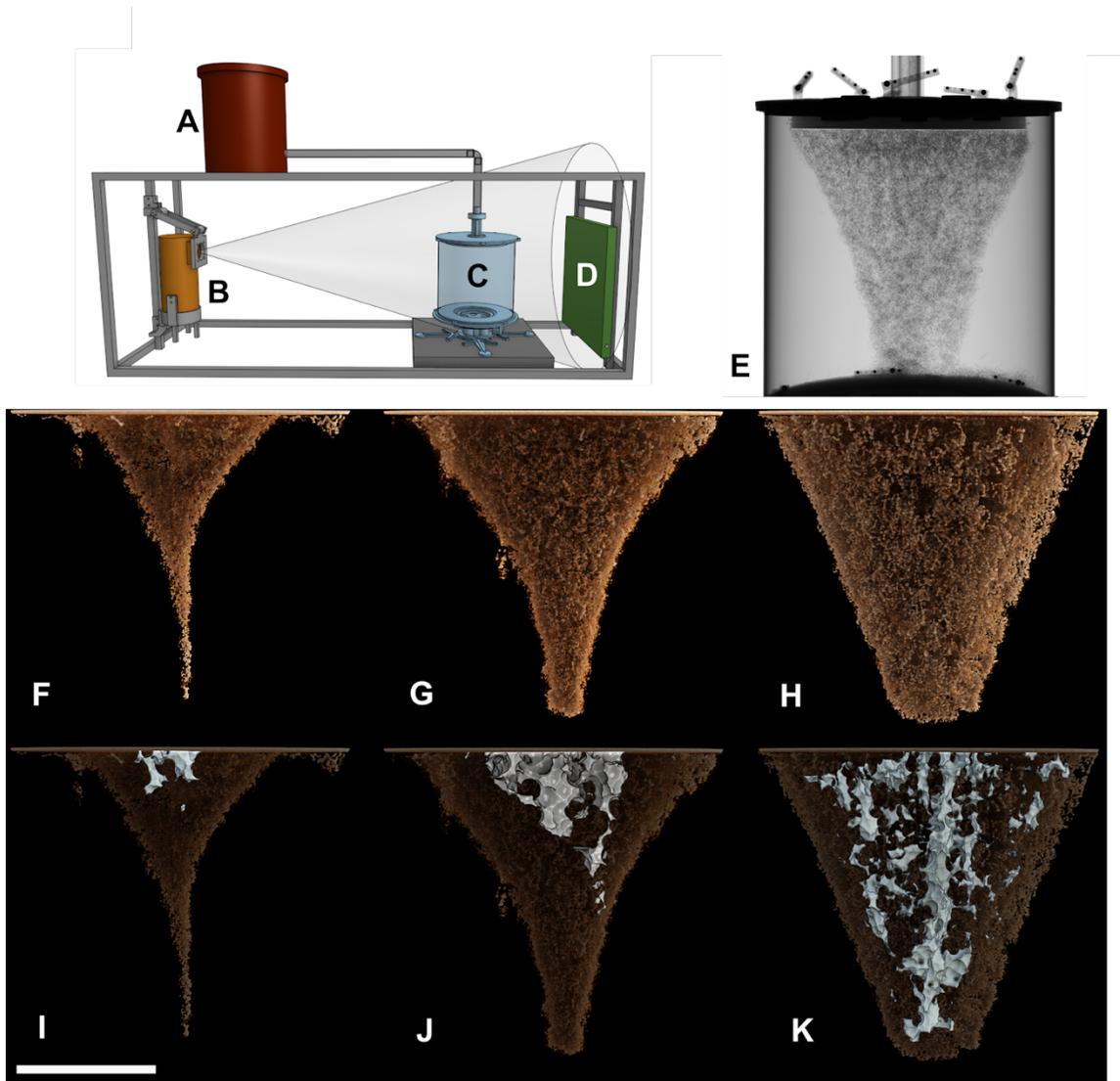

**Figure 2.** CT scanner equipment and reconstructions of bivouacs throughout the growth period, with visualizations of empty spaces. A–D, custom X-ray CT scanner diagram with A, ant collection container, B, X-ray source, C, imaging container on rotation stage, and D, X-ray detector (see Movies S2 and S4). Ants moved freely from the collection into the imaging container via a tube connecting A to C. E, example X-ray projection of the imaging container with bivouac. F–H, rendered view of bivouacs recorded at F, 20, G, 60, and H, 180 minutes; I–K their corresponding internal empty spaces. White scale bar below I represents 10 cm of length.

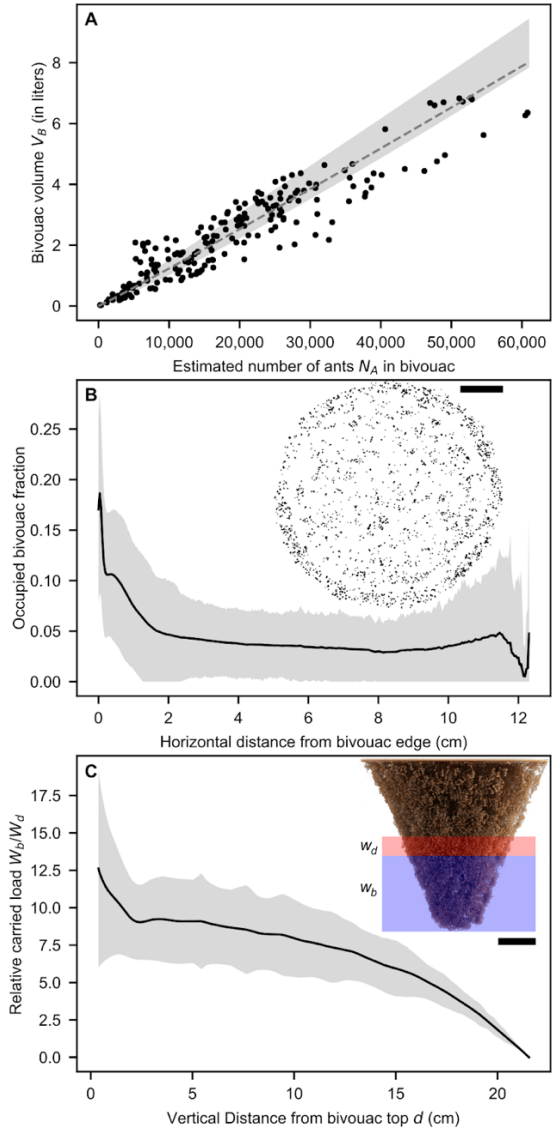

**Figure 3.** Structural properties of bivouacs. A, The scaling relationship of bivouac volume $V_B$ with the number of ants in the bivouac $N_A$ is positively allometric (dashed line, $V_B = -9.33 * N_A^{1.035}$), suggesting that bivouacs become less dense as they grow larger. The grey area gives the 95% confidence region for the power law fit parameters obtained via bootstrapping. B, bivouacs have a defined higher-density shell at the outer bivouac edge of approximately 1.4 cm thickness, and an internal volume approximately half as dense. The plot line gives the average of all horizontal slices in all bivouacs, the grey area indicates standard deviation. The inset shows an example horizontal slice through a bivouac, ant positions shown in black reveal areas of different density including the well-defined outer shell. C, relative load carried by ants throughout the bivouac volume. The load is calculated via a sliding window of thickness t = 0.76 cm, by dividing the integrated mass of the bivouac below this window ($W_b$) by the integrated mass within the window ($W_d$) (window width is not shown to scale). Subscript d refers to the distance between window center and bivouac top. The inset visualizes the calculation concept. The increase in relative load carriage at $d < 2$ cm is the result of the bivouac's upper region being spatially constrained by the experimental setup; in unconstrained bivouacs, this increase is absent (see Methods and Fig. S7). The scale bars in B,C represent 5 cm of length.

Supplementary Information for

# Anatomy of a superorganism – structure and growth dynamics of army ant bivouacs

Thomas Bochynek, Florian Schiffers, André Aichert, Oliver Cossairt, Simon Garnier, and Michael Rubenstein*

*Corresponding author: Michael Rubenstein
Email:  rubenstein@northwestern.edu

This PDF file includes:

Supplementary text
Figures S1 to S7
Tables S1 to S4
Legends for Movies S1 to S4

Other supplementary materials for this manuscript include the following:

Movies S1 to S4

**Detector pixel fault correction**

Our X-ray detector has approximately 100 permanently faulty pixels which either report full or zero intensity in every exposure. If not accounted for, these readings would result in significant artifacts in the reconstructions. We hence detected those pixels with a prior calibration measurement and corrected them using the inpainting function *inpaint* (with parameters *radius = 10, method = INPAINT_NS*) from *OpenCV* 3.4.1. All image processing was implemented in *Python* 3.7.3.

**3D volume reconstruction from X-ray projection data**

When X-rays pass through an object, their intensity is reduced. The quantity for this reduction, the attenuation coefficient $\mu$, is given by the Beer–Lambert attenuation law
$I = I_0 e^{-\int \mu(r) dr}$,
where *I* is the measured intensity, $I_0$ is the reference intensity without specimen, $\mu$ is the attenuation coefficient at spatial location $r$ and the integral is performed over the path of the X-ray.
Computed tomography seeks to, for a scanned specimen, calculate the X-ray attenuation coefficient for every spatial location (see (1) for a comprehensive introduction to computed tomography). This allows visualization of the internal structure of the specimen as a 3D volume. CT imaging uses a sequence of radiographic X-ray projections acquired by rotating the sample on a fixed rotation stage. These are normalized using a reference image (an exposed image without specimen) $I_0$ and a dark image (an unexposed image, i.e. a detector reading without X-rays present) $I_D$ by solving for the measured line-integral:
$\int \mu(r) dr = -\ln \frac{I - I_D}{I_O - I_D}$.

From the normalized projections, the attenuation coefficients can be calculated using image reconstruction techniques. We applied these methods via the existing CT reconstruction framework *CONRAD*(2) using the Feldkamp reconstruction algorithm (2). Image reconstruction was undertaken in *Python* 3.7.3.
Cone-Beam Computed Tomography requires accurate calibration of the geometrical setup. Because our experiment design required that the imaging container be removed after the conclusion of a replicate, we could not guarantee that the setup remains unchanged to the accuracies required for the image reconstruction. As a remedy, we attached fiducial markers (see Movie S2) to the imaging containers, which allowed for a post-hoc calibration and accurate image reconstruction (3, 4).

After tomographic reconstruction, we normalized the volumetric data similar to the Hounsfield unit transform to obtain $I = \frac{\mu - \mu_{cork}}{\mu_{cork} - \mu_{Air}}$, where $\mu$ is the reconstructed intensity, and $\mu_{Cork}$ and $\mu_{air}$ are extracted from each scan individually from several slices around the lid where only volumes of cork or air are present. After normalization, all tomographic reconstructions have the same unit and data can be compared quantitatively. Example reconstructions are shown in Fig. S1.

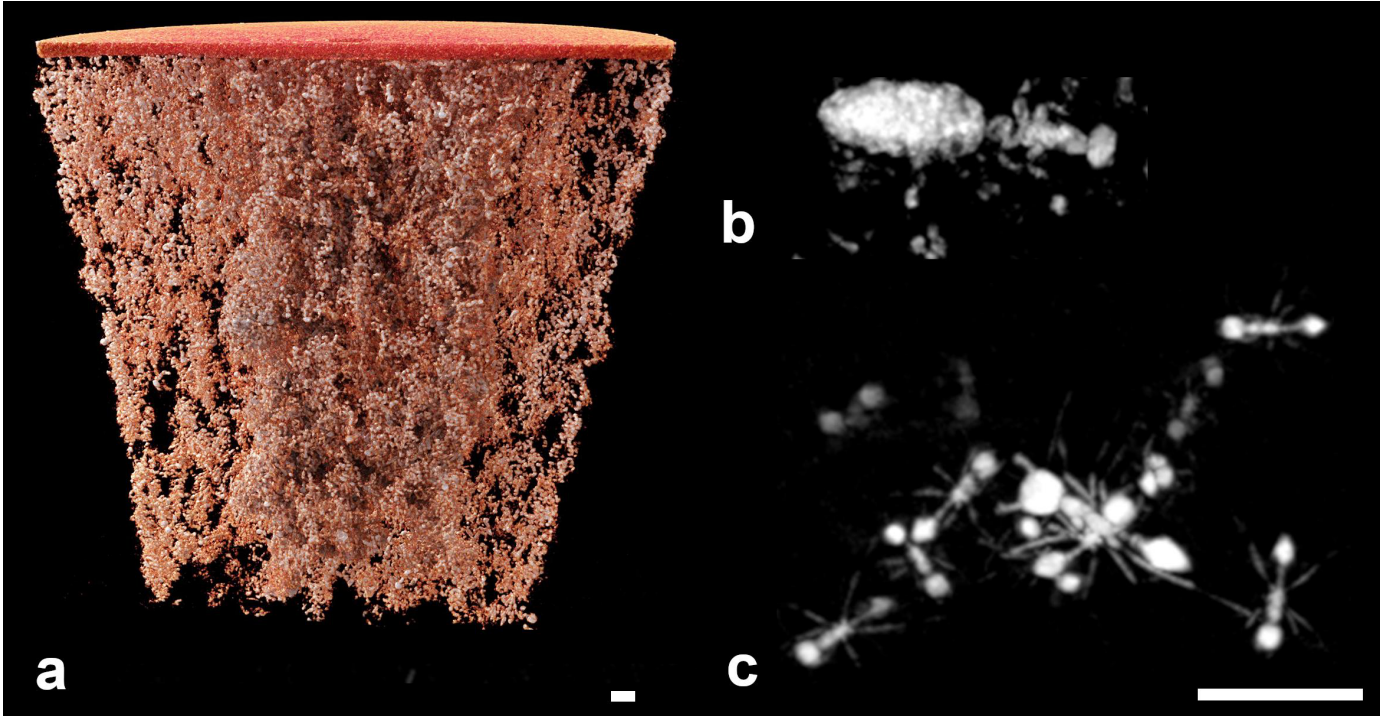

Fig. S1. Reconstructions of bivouac, queen, and workers. **a**, Sagittal cut through rendered 3D reconstruction of a bivouac shows bisected empty volumes. Visualization by Siemens Healthineers Cinematic Rendering Software. **b**, **c**, reconstruction of *Eciton burchellii* queen (**b**) and polymorphic workers (**c**). White scale bars represent 1 cm of length. Scale is identical in **b** and **c**.

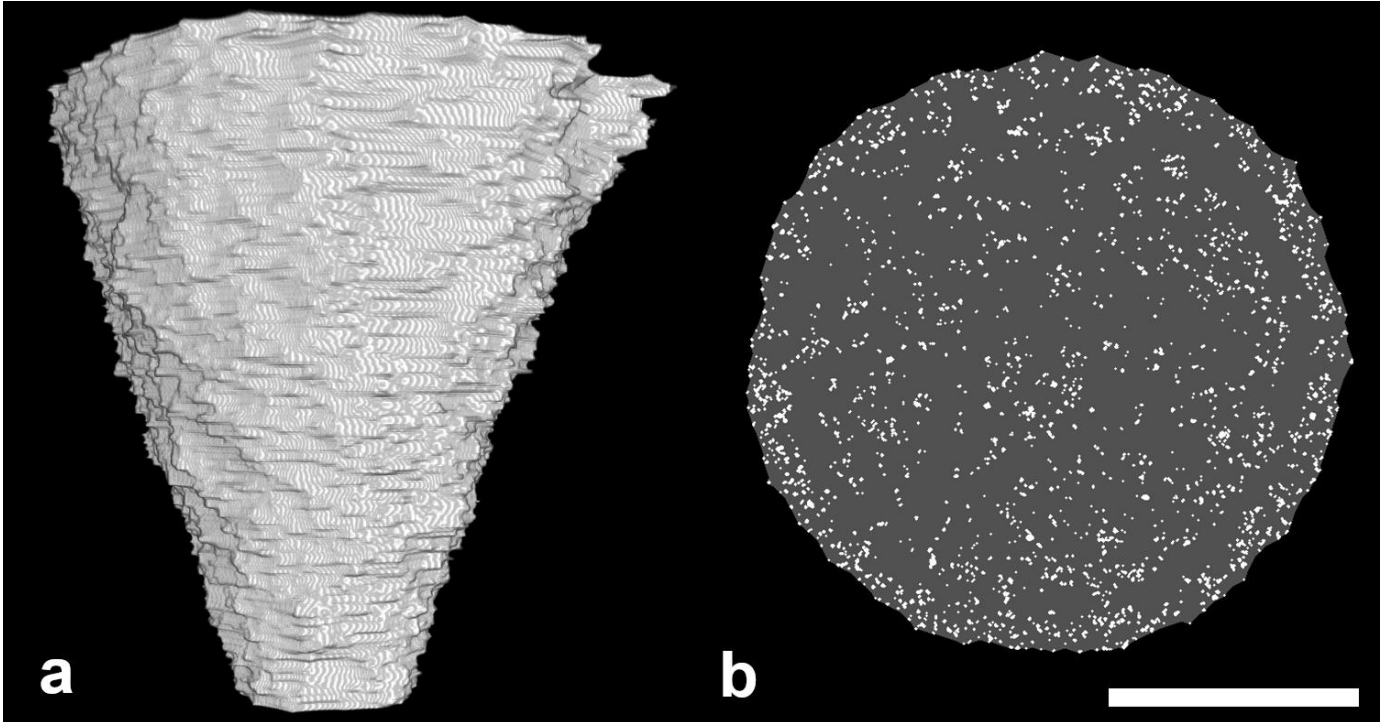

**Fig. S2**. Example of bivouac bounding hull. a, example of a bivouac bounding hull (rendered in Fiji/ImageJ 1.52p (5) 3DViewer plugin) and **b**, horizontal cross section of hull (grey) superimposed on ant position data (white), taken from the upper bivouac region. Scale bar represents 10 cm. Note that the hull-making algorithm attaches the hull perimeter directly to ant locations, over-representing ant density on the immediate hull periphery and creating the artefact shown in Fig. 3B.

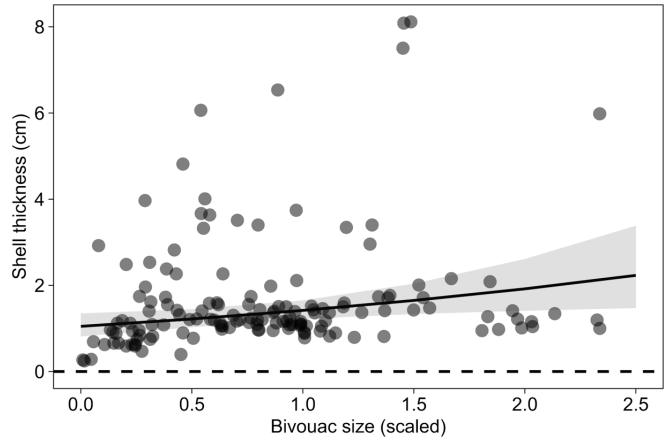

**Fig. S3.** Shell thickness as a function of bivouac size. Shell thickness was estimated as the breakpoint in a weighted segmented linear regression model on the occupancy ratio at different distances to the bivouac's exterior. The bivouac size is scaled using the root mean square of the data. The black line and greyed area correspond to the prediction of the linear mixed-effects model and to its confidence interval.

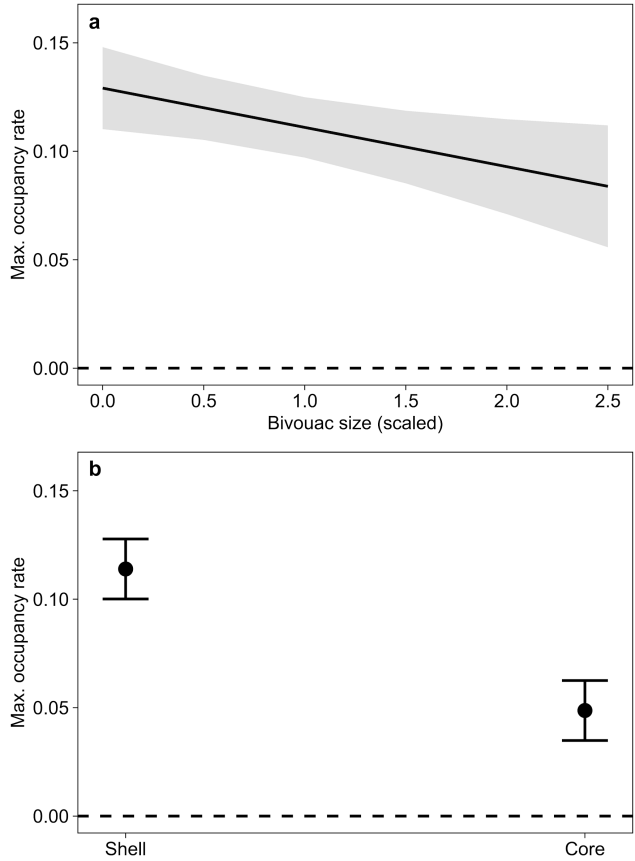

**Fig. S4.** Influence of bivouac size on maximum occupancy rate of each section. The maximum occupancy rates were estimated as the intercepts in a weighted segmented linear regression model on the occupancy ratio at different distances to the bivouac's exterior. The bivouac size is scaled using the root mean square of the data. **a,** Predicted effect of the bivouac size. The black line and greyed area correspond to the prediction of the linear mixed-effects model and to its confidence interval. **b,** Predicted effect of the section. The dots and confidence bars correspond to the predictions of the linear mixed-effects model for each section and to their confidence intervals.

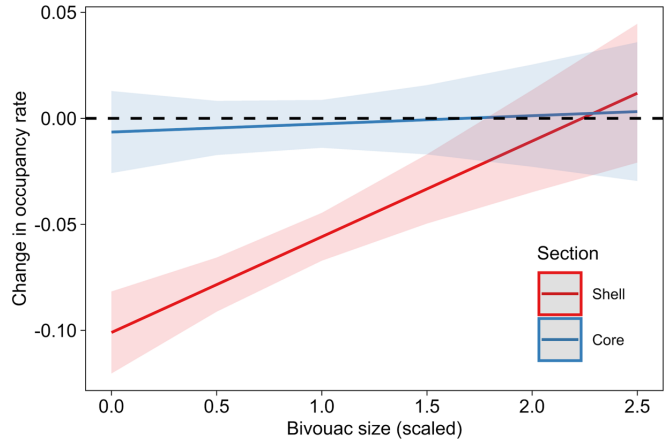

**Fig. S5.** Influence of bivouac size on changes in the occupancy rate of each section. Changes in the occupancy rates were estimated as the slopes in a weighted segmented linear regression model on the occupancy ratio at different distances to the bivouac's exterior. The bivouac size is scaled using the root mean square of the data. The red and blue lines and the transparent red and blue areas correspond to the predictions of the linear mixed-effects model and to their confidence intervals.

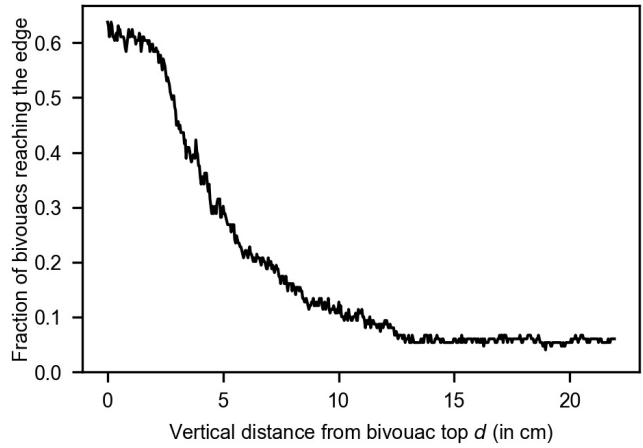

**Fig. S6.** Fraction of bivouacs that reach or exceed the edge of the cork disk. The fraction of these bivouacs declines sharply within the upper $d = 5$ cm of the bivouac. This range of decline coincides with the region of increased load carriage shown in Fig. 3C.

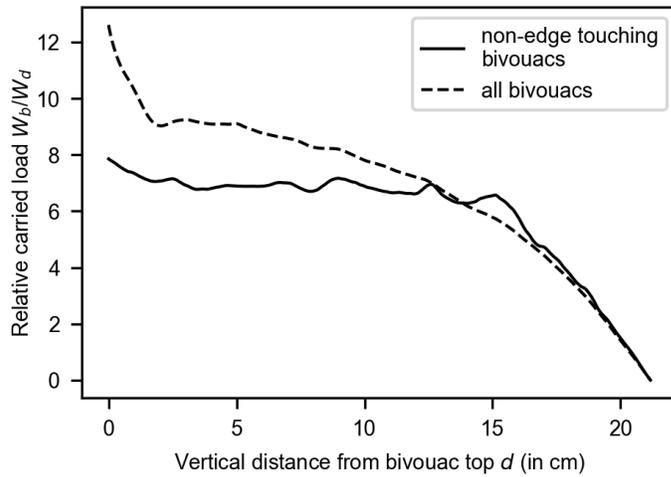

**Fig. S7.** Comparison of carried load for non-edge touching bivouacs compared to that of all imaged bivouacs. In those bivouacs that never touch the edge of the cork disk ($N = 25$), the carried load increase at the bivouac's upper region seen in the average of all bivouacs is absent. This suggests that the increase is a result of the spatial constraint exerted on larger bivouacs by the limited available attachment space provided by the cork disk.

**Table S1.** Bivouac occupancy persistence data. Bivouac occupancy is transient: of 1,189,965,249 voxels occupied at least once, only 277,181,652 (i.e. 23.3%) remained occupied until the end of the experiment.

| Experiment | Voxels occupied at least once | Voxels occupied until end after first occupancy |
|---|---|---|
| 1 | 25,844,024 | 1,092,474 |
| 2 | 15,967,238 | 2,014,838 |
| 3 | 8,064,226 | 1,363,259 |
| 4 | 3,940,558 | 1,585,502 |
| 5 | 41,309,472 | 4,861,621 |
| 6 | 45,953,242 | 7,149,371 |
| 7 | 35,304,215 | 9,333,605 |
| 8 | 44,887,517 | 9,628,147 |
| 9 | 57,759,492 | 4,985,784 |
| 10 | 39,820,511 | 10,392,175 |
| 11 | 40,294,650 | 6,069,854 |
| 12 | 44,697,926 | 10,167,303 |
| 13 | 69,476,436 | 14,266,335 |
| 14 | 86,001,101 | 35,811,399 |
| 15 | 57,737,495 | 24,067,989 |
| 16 | 100,988,591 | 20,870,535 |
| 17 | 98,023,380 | 18,428,350 |
| 18 | 104,946,311 | 8,398,760 |
| 19 | 62,547,321 | 9,054,436 |
| 20 | 26,864,484 | 11,598,870 |
| 21 | 40,587,521 | 33,565,073 |
| 22 | 61,881,828 | 13,208,366 |
| 23 | 18,402,435 | 4,597,264 |
| 24 | 44,854,829 | 7,279,779 |
| 25 | 13,810,446 | 7,390,563 |
| Sum | 1,189,965,249 | 277,181,652 |

**Table S2.** Effect of bivouac size on bivouac shell thickness. Statistics and coefficients of the linear mixed-effects model. Bivouac shell thickness was positively correlated with bivouac size ($p = 0.015$).

| Predictors | log(breakpoint) | | |
|---|---|---|---|
| | Estimates | CI | p |
| (Intercept) | 0.05 | -0.20 – 0.30 | 0.717 |
| Scaled bivouac size | 0.30 | 0.07 – 0.54 | 0.015 |
| **Random Effects** | | | |
| σ2 | 0.26 | | |
| τ00 colony_id:bivouac_id | 0.09 | | |
| ICC | 0.26 | | |
| N colony_id | 5 | | |
| N bivouac_id | 7 | | |
| Observations | 147 | | |
| Marginal R2 / Conditional R2 | 0.073 / 0.315 | | |

**Table S3.** Effect of bivouac size on the maximum occupancy ratio of each section (core vs shell). Statistics and coefficients of the linear mixed-effects model. The maximum occupancy ratio was significantly greater in the shell than the core ($p < 0.001$) and overall negatively correlated to the bivouac size ($p = 0.019$).

| Predictors | Estimates | CI | p |
|---|---|---|---|
| (Intercept) | 0.13 | 0.11 – 0.15 | <0.001 |
| section [Core] | -0.07 | -0.07 – -0.06 | <0.001 |
| scaled bivouac size | -0.02 | -0.03 – 0.00 | -0.019 |
| **Random Effects** | | | |
| $\sigma^2$ | 0.00 | | |
| $\tau_{00}$ colony_id:bivouac_id | 0.00 | | |
| ICC | 0.49 | | |
| N colony_id | 5 | | |
| N bivouac_id | 7 | | |
| Observations | 294 | | |
| Marginal R2 / Conditional R2 | 0.358 / 0.671 | | |

**Table S4**. Effect of bivouac size on occupancy ratio changes inside each section (core vs shell). Statistics and coefficients of the linear mixed-effects model. The occupancy ratio decreases significantly faster in the shell than in the core for small bivouac sizes but not for large bivouac sizes (interaction section x bivouac size, $p < 0.001$).

|  | slope | | |
| --- | --- | --- | --- |
| Predictors | Estimates | CI | p |
| (Intercept) | -0.10 | -0.12 – -0.08 | <0.001 |
| section [Core] | 0.09 | 0.07 – 0.12 | <0.001 |
| Scaled bivouac size | 0.05 | 0.03 – 0.06 | <0.001 |
| section [Core] * scaled bivouac size | -0.04 | -0.06 – -0.02 | <0.001 |
| Random Effects | | | |
| σ2 | 0.00 | | |
| τ00 colony_id:bivouac_id | 0.00 | | |
| ICC | 0.12 | | |
| N colony_id | 5 | | |
| N bivouac_id | 7 | | |
| Observations | 294 | | |
| Marginal R2 / Conditional R2 | 0.301 / 0.382 | | |

**Legends for Movies S1 to S4**

Movie S1: https://youtu.be/G5qiROIMrBo. Rendering of bivouac internal structure

Movie S2: https://youtu.be/WrNoY4MgAw0. Projection sequence and on-line calibration

Movie S3: https://youtu.be/UfX_ldpky4E. Bivouac renders time series

Movie S4: https://youtu.be/LgB9ElBNOT4. Colony collection and CT scanner setup